\documentclass[twocolumn]{aastex631}

\newcommand{\yggdrasil}{\textsc{yggdrasil}}

\newcommand{\msun}{M$_\odot$}
\newcommand{\lsun}{L$_\odot$}
\newcommand{\sourcextractor}{{\tt SourceXtractor++}}
\newcommand{\ml}{$\Upsilon_*$}
\newcommand{\mstar}{M$_*$}
\newcommand{\per}{$^{-1}$}

\newcommand{\citepJD}{M.~J. Jim\'enez-Donaire et al. in prep.}
\newcommand{\change}{}

\shorttitle{Star Clusters in M\,82 Starburst}
\shortauthors{Levy et al.}

\begin{document}

\title{JWST Observations of Starbursts: Massive Star Clusters in the Central Starburst of M\,82}

\correspondingauthor{Rebecca C. Levy}
\email{rebeccalevy@arizona.edu}


\newcommand{\AAPF}{\altaffiliation{NSF Astronomy and Astrophysics Postdoctoral Fellow}}

\newcommand{\Arizona}{\affiliation{Steward Observatory, University of Arizona, Tucson, AZ 85721, USA}}

\newcommand{\Maryland}{\affiliation{Department of Astronomy, University of Maryland, College Park, MD 20742, USA}}

\newcommand{\JSI}{\affiliation{Joint Space-Science Institute, University of Maryland, College Park, MD 20742, USA}}

\newcommand{\INA}{\affiliation{Instituto Nacional de Astrof\'isica, \'Optica y Electr\'onica, Luis Enrique Erro 1, Tonantzintla, 72840 Puebla, Mexico}}

\newcommand{\BUAP}
{\affiliation{Benem\'erita Universidad Aut\'onoma de Puebla, Av. San Manuel, 72000 Puebla, Mexico}}

\newcommand{\STScI}{\affiliation{Space Telescope Science Institute, 3700 San Martin Drive, Baltimore, MD 21218, USA}}

\newcommand{\MPIA}{\affiliation{Max Planck Institut for Astronomy, Konigstuhl 17, 69117 Heidelberg, Germany}}

\newcommand{\KU}{\affiliation{Department of Physics and Astronomy, University of Kansas, 1251 Wescoe Hall Drive, Lawrence, KS 66045, USA}}

\newcommand{\UWyoming}{\affiliation{Department of Physics and Astronomy, University of Wyoming, Laramie, WY 82071, USA}}

\newcommand{\Leiden}{\affiliation{Leiden Observatory, Leiden University, P.O.~Box 9513, 2300~RA~Leiden, The Netherlands}}

\newcommand{\UGent}{\affiliation{Sterrenkundig Observatorium, Ghent University, Krijgslaan 281 - S9, B-9000 Gent, Belgium}}

\newcommand{\France}{\affiliation{Coll\`ege de France, 11 Pl. Marcelin Berthelot, 75231 Paris, France}}

\newcommand{\ParisObs}{\affiliation{Observatoire de Paris, 61 avenue de l’Observatoire, 75014 Paris, France}}

\newcommand{\ESOST}{\affiliation{European Space Agency, c/o STScI, 3700 San Martin Drive, Baltimore, MD 21218, USA}}

\newcommand{\ITA}{\affiliation{Universit\"{a}t Heidelberg, Zentrum f\"{u}r Astronomie, Institut f\"{u}r Theoretische Astrophysik, Albert-Ueberle-Str. 2, D-69120 Heidelberg, Germany}}

\newcommand{\IWR}{\affiliation{Universit\"{a}t Heidelberg, Interdisziplin\"{a}res Zentrum f\"{u}r Wissenschaftliches Rechnen, Im Neuenheimer Feld 205, D-69120 Heidelberg, Germany}}

\newcommand{\SOFIA}{\affiliation{Stratospheric Observatory for Infrared Astronomy, NASA Ames Research Center, Mail Stop 204-14, Moffett Field, CA 94035, USA}}

\newcommand{\JPL}{\affiliation{Jet Propulsion Laboratory, California Institute of Technology, 4800 Oak Grove Dr., Pasadena, CA 91109, USA}}

\newcommand{\UdeC}{\affiliation{Departamento de Astronom\'ia, Universidad de Concepci\'on, Barrio Universitario, Concepci\'on, Chile}}

\newcommand{\NMMT}{\affiliation{New Mexico Institute of Mining and Technology, 801 Leroy Place, Socorro, NM 87801, USA}}

\newcommand{\NRAOSocorro}{\affiliation{National Radio Astronomy Observatory, P.O. Box O, 1003 Lopezville Road, Socorro, NM 87801, USA}}

\newcommand{\OSU}{\affiliation{Department of Astronomy, The Ohio State University, Columbus, OH 43210, USA}}

\newcommand{\UGR}{\affiliation{Dept. F\'isica Te\'orica y del Cosmos, Universidad de Granada, 18071, Granada, Spain }}

\newcommand{\OAN}{\affiliation{Observatorio Astronómico Nacional (IGN), C/Alfonso XII, 3, E-28014 Madrid, Spain}}

\newcommand{\YS}{\affiliation{Centro de Desarrollos Tecnológicos, Observatorio de Yebes (IGN), 19141 Yebes, Guadalajara, Spain}}

\newcommand{\Swinburne}{\affiliation{Centre for Astrophysics and Supercomputing, Swinburne University of Technology, Hawthorn, VIC 3122, Australia}}

\newcommand{\ASTROTD}{\affiliation{ARC Centre of Excellence for All Sky Astrophysics in 3 Dimensions (ASTRO 3D)}}


\author[0000-0003-2508-2586]{Rebecca C. Levy}
\AAPF
\Arizona

\author[0000-0002-5480-5686]{Alberto D. Bolatto}
\Maryland
\JSI

\author[0000-0002-4677-0516]{Divakara Mayya}
\INA

\author[0000-0002-1046-1500]{Bolivia Cuevas-Otahola}
\BUAP

\author[0000-0003-1356-1096]{Elizabeth Tarantino}
\STScI

\author[0000-0003-4850-9589]{Martha L. Boyer}
\STScI

\author[0000-0002-3952-8588]{Leindert A. Boogaard}
\MPIA

\author[0000-0002-5666-7782]{Torsten B\"oker}
\ESOST

\author[0000-0002-9511-1330]{Serena A. Cronin}
\Maryland

\author[0000-0002-5782-9093]{Daniel~A.~Dale}
\UWyoming

\author[0009-0004-5807-9142]{Keaton Donaghue}
\KU

\author[0000-0001-6527-6954]{Kimberly L.~Emig}
\France
\ParisObs

\author[0000-0003-0645-5260]{Deanne B. Fisher}
\Swinburne
\ASTROTD

\author[0000-0001-6708-1317]{Simon C.~O.~Glover}
\ITA

\author[0000-0002-2775-0595]{Rodrigo Herrera-Camus}
\UdeC

\author[0000-0002-9165-8080]{Mar\'ia J. Jim\'enez-Donaire}
\OAN
\YS

\author[0000-0002-0560-3172]{Ralf S. Klessen}
\ITA
\IWR

\author[0000-0003-4023-8657]{Laura Lenki\'{c}}
\SOFIA
\JPL

\author[0000-0002-2545-1700]{Adam K. Leroy}
\OSU

\author[0000-0001-9419-6355]{Ilse De Looze}
\UGent

\author[0000-0001-9436-9471]{David S. Meier}
\NMMT
\NRAOSocorro

\author[0000-0001-8782-1992]{Elisabeth A.~C. Mills}
\KU

\author[0000-0001-8224-1956]{Juergen Ott}
\NRAOSocorro

\author[0000-0003-1682-1148]{M\'onica Rela\~no}
\UGR

\author[0000-0002-3158-6820]{Sylvain Veilleux}
\Maryland
\JSI

\author[0000-0002-5877-379X]{Vicente Villanueva}
\UdeC

\author[0000-0003-4793-7880]{Fabian Walter}
\MPIA

\author[0000-0001-5434-5942]{Paul P.~van der Werf}
\Leiden

\begin{abstract}
We present a near infrared (NIR) candidate star cluster catalog for the central kiloparsec of M\,82 based on new JWST NIRCam images. We identify star cluster candidates using the F250M filter, finding 1357 star cluster candidates with stellar masses $>10^4$~M$_\odot$. Compared to previous optical catalogs, nearly all (87\%) of the candidates we identify are new. The star cluster candidates have a median intrinsic cluster radius of $\approx$1~pc and have stellar masses up to $10^6$~M$_\odot$. By comparing the color-color diagram to dust-free \textsc{yggdrasil} stellar population models, we estimate that the star cluster candidates have A$_{\rm V}\sim3-24$~mag, corresponding to A$_{\rm 2.5\mu m}\sim0.3-2.1$~mag. There is still appreciable dust extinction towards these clusters into the NIR. We measure the stellar masses of the star cluster candidates, assuming ages of 0 and 8~Myr. The slope of the resulting cluster mass function is $\beta=1.9\pm0.2$, in excellent agreement with studies of star clusters in other galaxies.

\end{abstract}


\section{Introduction}
\label{sec:intro}

The cycle of star formation is a key driver of the evolution of galaxies. At high surface densities of star formation, the majority of stars form in gravitationally bound clusters \citep[e.g.,][]{Kruijssen2012,Krause2020}. Various phases of star cluster formation inject energy and momentum (i.e., feedback) into the surrounding interstellar medium (ISM). Recently, it has become clear that pre-supernovae feedback can be a dominant mechanism to clear the natal gas from a star cluster \citep[e.g.,][]{Chevance2022,Farias2024}. However, the earliest phases of star cluster formation and evolution are enshrouded in dust and gas, making them difficult to observe. Multiwavelength observations of star clusters are therefore necessary to probe the full cycle of star cluster evolution.

In regions of high molecular gas surface densities, the star formation rates (SFRs) are higher than expected for a simple scaling of the Kennicutt-Schmidt relation in more typical conditions \citep[e.g.,][]{Kennicutt2012}. This suggests that the process of star formation in starburst regions is physically different than other regimes. One consequence, seen in both observations and models, is that the fraction of stars formed in clusters increases with the surface density of star formation (e.g., \citealt{Goddard2010,Adamo2011,Silva-Villa2011,Kruijssen2012,Leroy2018}, though see \citealt{Cook2023}). Moreover, this high activity mode tends to form ``super" star clusters, massive (M$_*\gtrsim10^5$~\msun) and compact ($r\lesssim1$~pc) bound groups of stars \citep[e.g.,][]{PortegiesZwart2010}.


The  small, irregular galaxy M\,82 (NGC\,3034; M$_{\rm dyn}\approx10^{10}$~\msun, $R_{\rm dyn}\approx4$~kpc; \citealt{Greco2012,Dale2023}) is the archetypal starburst in the local Universe. Located in the M\,81 group at a distance of 3.6~Mpc \citep{Freedman1994}, the starburst in M\,82 is driven by the tidal interactions in the group \citep[e.g.,][]{Yun1994,ForsterSchreiber2003}. The starburst region in M\,82 roughly extends over the central kiloparsec. M\,82 has experienced two recent bursts of star formation \citep{ForsterSchreiber2003}. The first occurred between $8-15$~Myr ago, peaked at a SFR of 160~\msun~yr\per, and was driven primarily by tidal interactions. The second occurred $4-6$~Myr ago, peaked at a SFR of 40~\msun~yr\per, and was driven by a bar inflow \citep[the formation of which likely resulted from the interaction, e.g.,][]{ForsterSchreiber2003}. M\,82 has a ``quiescent" total infrared luminosity-derived SFR~$\simeq 12$~\msun~yr\per\ \citep{Herrera-Camus2018a,Herrera-Camus2018b}, which traces the SFR over the last 100~Myr \citep[e.g.,][]{Calzetti2013}. A result of the starburst phases in the recent past is the massive multiphase outflow of material from the central starburst region \citep[see e.g.,][and references therein]{Bolatto2024} --- a defining feature of M\,82.

A number of massive star clusters have been identified and characterized in the central starburst of M\,82 using several facilities, including with the Hubble Space Telescope (HST) using optical \citep{O'Connell1995,Melo2005,Mayya2008} and NIR \citep{McCrady2003} images, optical-infrared spectroscopy \citep{McCrady2003,McCrady2005,McCrady2007,Westmoquette2014}, mid-infrared (MIR) imaging \citep{Gandhi2011}, and Submillimeter Array (SMA) observations (\citepJD). The HST-based cluster catalog of \citet{Mayya2008} is the most complete accounting of the optical star clusters in M\,82. Using HST F435W, F555W, and F814W images, they identified 260 star clusters within the central 900~pc and 363 star clusters outside of the central starburst. The pixel scale of these images is 0.05\arcsec\ (0.88~pc). In order to reject compact ``starlike" objects from their catalog, \citet{Mayya2008} require that the candidates have a minimum (maximum) diameter of 3 (30) pixels and an area of $>50$~pixels. Therefore, the clusters in their catalog have Gaussian radii $>1.3$~pc and areas of $\gtrsim40~{\rm pc}^2$. Their clusters tend to have A$_{\rm V}\lesssim6$, and their cluster mass functions have a power-law slope of 1.8 for clusters in the starburst, assuming an age of 8~Myr.

The central region has a patchy distribution of dust with extinction in excess of A$_{\rm V}=30$~mag for the NIR-detected star clusters \citep{ForsterSchreiber2003}. As young star clusters are compact (radii~$\lesssim1$~pc; e.g., \citealt{PortegiesZwart2010,Krumholz2019,Brown2021,Levy2022}) and tend to resolve into smaller sub-structures \citep[e.g.,][]{Levy2021,Levy2022}, high spatial resolution observations are needed to obtain an accurate census of star cluster populations. JWST offers parsec-scale spatial resolutions for nearby galaxies such as M\,82 at unprecedented sensitivities, which provides a unique opportunity to identify and study clusters over a range of evolutionary stages from embedded to more exposed. The reduced extinction at NIR wavelengths as compared to traditional optical wavelengths allows us to obtain a more complete catalog of all star clusters. Both NIRCam and MIRI onboard JWST have already proven to be powerful tools to detect and characterize star clusters in nearby galaxies \citep[e.g.,][]{Linden2023,Rodriguez2023,Schinnerer2023,Sun2024}.

In this Letter, we construct and present the JWST NIRCam star cluster catalog in the central region of the M\,82 starburst. 
We briefly present the observations and data processing in Section~\ref{sec:obs}. In Section~\ref{sec:prop}, we describe the NIRCam star cluster catalog. We discuss the overlaps with star cluster catalogs at other wavelengths, color-color diagrams, and the cluster mass function in Section~\ref{sec:disc}. We summarize our results in Section~\ref{sec:summary}.

\section{Observations and Data Reduction}
\label{sec:obs}

We identify clusters using NIRCam data obtained as part of JWST Cycle 1 GO program \#1701 (P.I. Alberto Bolatto). We direct the reader to \citet{Bolatto2024} for full details of the observations and data reduction. Briefly, we use NIRCam \citep{Rieke2023} \texttt{SUB640} observations of the central 50\arcsec\ ($\approx$880~pc) of M\,82\footnote{Automatic pipeline processed MAST mosaics can be found under \url{http://dx.doi.org/10.17909/cwtn-nh63}.}. A three color image of the continuum filters from Program \#1701 (F140M, F250M, and F360M) is shown in Figure \ref{fig:multicolor}, which was produced with \texttt{multicolorfits} \citep{multicolorfits}. The left panel of each inset pair shows the same filters over four cluster-rich 50~pc~$\times$~50~pc regions.

Our JWST program observed in three ``continuum" filters: F140M, F250M, and F360M to enable a broad range of science goals (see \citealt{Bolatto2024} for more details). While the F140M filter has higher angular resolution among these three filters, it suffers the most from dust extinction meaning some clusters will be missed (e.g., Figure \ref{fig:multicolor} and Figure 2 of \citealt{Bolatto2024}). F360M, on the other hand, has the lowest angular resolution and the continuum is contaminated by the 3.3~\micron\ polycyclic aromatic hydrocarbon (PAH) feature as well as a water-ice absorption feature, which may be particularly prominent in the young star clusters (see \citealt{Sandstrom2023} and \citealt{Bolatto2024} for a more in-depth discussion). Based on earlier spectroscopy of M\,82 \citep{Sturm2000,ForsterSchreiber2001}, we expect the F250M filter to be largely free from line emission and to provide a clean estimate of the stellar continuum. Therefore, F250M is the ideal filter in our suite of NIRCam observations with which to identify NIR-bright star clusters. The F250M filter has a point-spread-function (PSF) full-width-half-maximum (FWHM) of 0.085\arcsec\ (1.5~pc) and a pixel scale of 0.042\arcsec\ (0.7~pc).

\subsection{Ancillary Data}
\label{ssec:ancdata}

\begin{figure*}
    \centering
    \includegraphics[width=\textwidth]{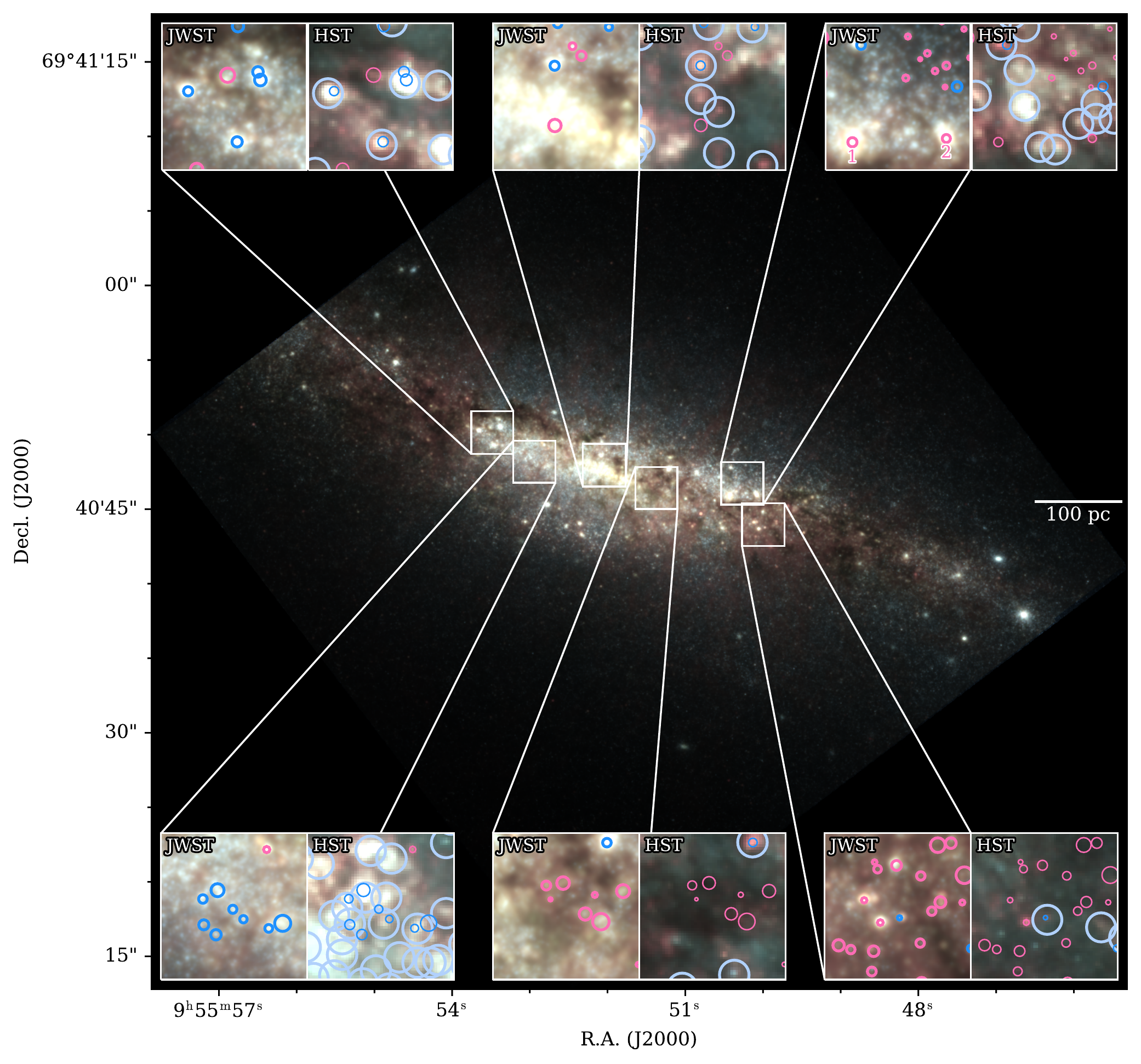}
    \caption{The central 870~pc of M\,82 seen with NIRCam showing F140M (blue), F250M (green), and F360M (red) on an asinh scale. \change{Insets show zoom-ins to 50~pc~$\times$~50~pc regions around some representative massive star cluster candidates.} The circles indicate the massive clusters identified in the NIRCam image, where the radius of the circle is the fitted (deconvolved) radius (see Section \ref{sec:prop} and Table \ref{tab:catalog}). Blue (pink) circles show clusters which do (do not) overlap with the HST catalog \citep[][see Section~\ref{ssec:comp_other_cats}]{Mayya2008}. Clusters 1 and 2 are labeled (see Sections~\ref{ssec:radii} and \ref{ssec:comp_other_cats}). The right insets show HST three-color images at the same locations (blue: F435W, green: F555W, red: F814W). The light blue circles in the HST insets show the star clusters identified by \citet{Mayya2008} in these selected regions using the same HST data, where the radius of each circle is 5~pc. The pink and blue circles are the same as in the left panels.}
    \label{fig:multicolor}
\end{figure*}

For comparison, in Figure \ref{fig:multicolor} we show data from HST in the right panel of each zoom-in pair. The selected filters --- F435W, F555W, F814W --- were downloaded from MAST\footnote{Automatic pipeline processed MAST mosaics can be found under \url{http://dx.doi.org/10.17909/mj38-1s44}.} and are the same as those used by \citet{Mayya2008} to catalog the star clusters across M\,82. These images have pixel scales of 0.05\arcsec. It is known that some HST images have astrometric offsets $\lesssim$1\arcsec\ with respect to e.g., interferometric images \citep[e.g.,][]{Leroy2018,Levy2021}. Because \citet{Bolatto2024} found that these JWST NIRCam images are well aligned to a radio catalog of supernova remnants, we assume that the absolute astrometry of the NIRCam images is correct. To align the HST images to the JWST NIRCam data, we measured offsets to shift the HST astrometry by ($\alpha$,$\delta$) = (-1.52\arcsec, -0.67\arcsec) so that bright clusters visible in both images were cospatial. The light blue circles in the HST insets of Figure~\ref{fig:multicolor} show the cataloged star clusters in the central region from \citet{Mayya2008}.

\section{Star Cluster Properties}
\label{sec:prop}

\subsection{Cluster Identification}
\label{ssec:catalog}

Star cluster candidates were identified from the F250M image using \sourcextractor\footnote{\url{https://github.com/astrorama/SourceXtractorPlusPlus}} \citep{sourcextractor++1,sourcextractor++2}. We started from the default configuration file, with the following changes. The detection threshold above the background ({\tt detection-threshold}; i.e., the signal-to-noise ratio relative to the measured local background) was increased to 10 due to the presence of numerous bright point sources (see the insets in Figure~\ref{fig:multicolor}). The minimum number of contiguous pixels above the detection threshold to qualify as a detection ({\tt detection-minimum-area}) was increased to 9~pixels to remove definite point sources\footnote{We note that this will also remove artifacts from hot pixels.} from the initial catalog. We also enable source cleaning ({\tt use-cleaning=1}) which removes false detections near bright objects in the preliminary catalog (though, as discussed below, this step removes very few sources). With these settings, \sourcextractor\ identified 2472 candidates.

\sourcextractor\ sometimes defines two clusters that significantly overlap. From visual inspection, we determine that such cases are better described as a single cluster. The fractional overlap of each candidate with every other candidate is measured using their center positions and a radius equal to the PSF FWHM. We flag those candidates with $>$75\% of their areas overlapping. We replace the coordinates of the overlapping clusters with the median of their coordinates determined by \sourcextractor. As described above, \sourcextractor\ tends to find false sources near bright objects. Although the source cleaning step should remove these, we found that, in practice, very few clusters were excluded during this step. To account for this, we remove candidates with slightly less overlap but where the weaker source does not appear to be cluster-like based on visual inspection. We determine the fractional overlap with a radius equal to twice the PSF FWHM, and we flag those candidates with $>$50\% of their areas overlapping. We extract the flux in a circular aperture of the same radius and keep only the brightest overlapping cluster. \sourcextractor\ also tends to pick up false sources along the edges of the image. We remove these from the preliminary catalog by discarding sources within 10 pixels of the image edge. After these initial cuts to remove spurious sources, we are left with 1767 star cluster and stellar candidates. We will discuss further refinements to this catalog in Section \ref{ssec:finalcatalog}.

\subsection{Radial Profiles and Radii Measurements}
\label{ssec:radii}

\begin{figure}[t]
    \centering
    \includegraphics[width=\columnwidth]{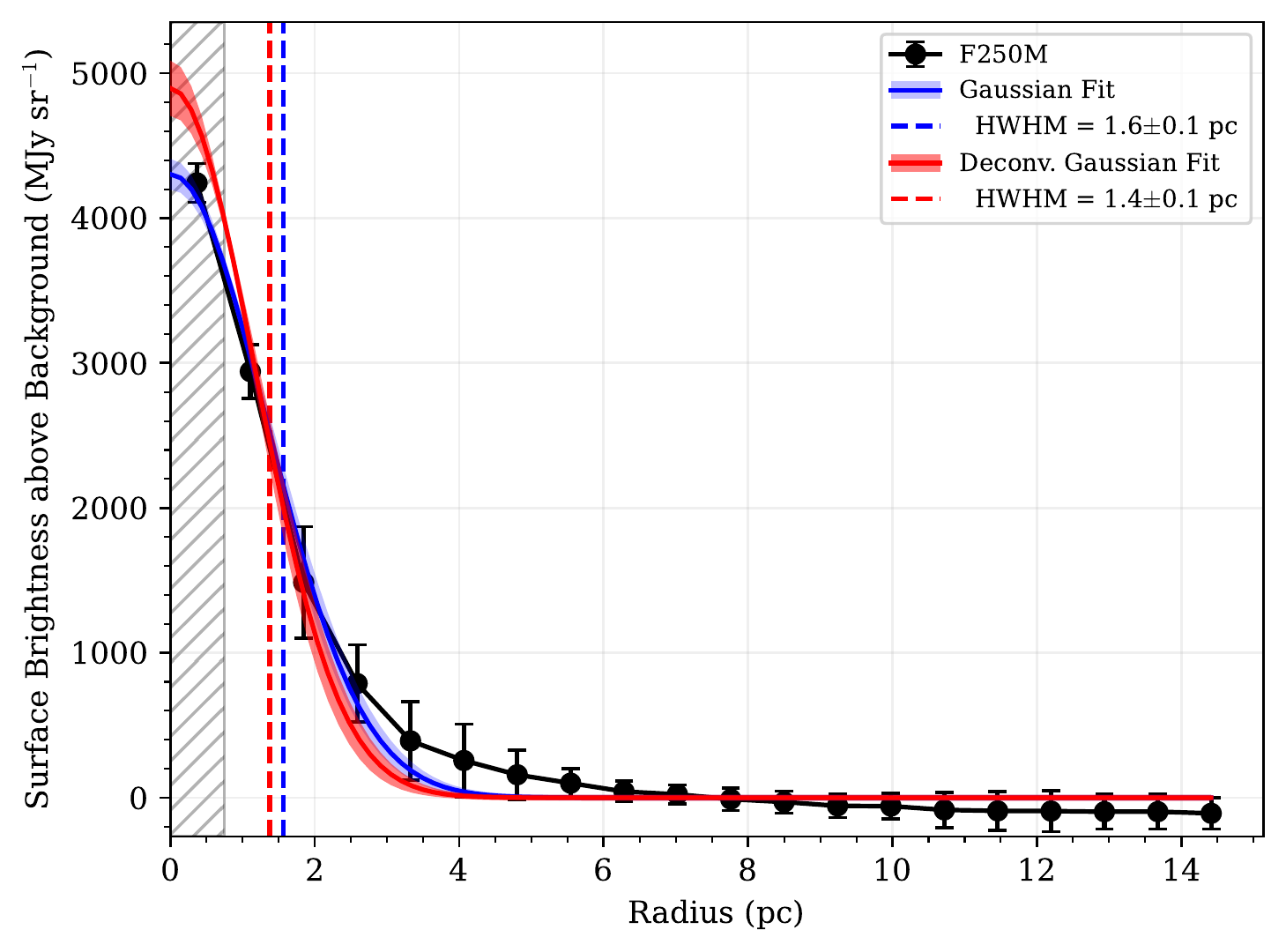}
    \caption{An example radial profile and Gaussian fit, shown here for Cluster 2 (marked in the upper right JWST inset in Figure \ref{fig:multicolor}). The black points show the radial profile extracted from the F250M image, where the error bars reflect the standard deviation in each ring. The blue curve and shaded region show the Gaussian fit to the radial profile and the uncertainties. The vertical blue dashed line indicates the HWHM of the fitted Gaussian. The gray vertical shaded region shows the HWHM of the F250M PSF. The red curve and shaded region show the deconvolved Gaussian fit, and the vertical red dashed line shows the HWHM of the deconvolved Gaussian (reported in Table~\ref{tab:catalog}).}
    \label{fig:radial_profile}
\end{figure}

We measure the size of each star cluster candidate by constructing a radial profile. Our methodology closely follows that of \citet[][see their Section 4.3]{Levy2022}. First we select a square region around each cluster that is 10$\times$ the PSF FWHM on each side (0.85\arcsec~$\approx$~15~pc). We then mask out other cluster candidates in this region (based on the positions in the preliminary star cluster candidate catalog) using a circle with a diameter equal to twice the PSF FWHM. This mitigates the effects of other bright objects from affecting the radial profile of the desired source. We extract the radial profile for each candidate in circular annuli centered on the coordinates determined by \sourcextractor\ and with a width equal to 1 pixel. We take the median of the surface brightness in each ring, and the uncertainty is the standard deviation. An example for one cluster is shown in Figure \ref{fig:radial_profile}.

We then fit each radial profile with a Gaussian function plus a constant offset (e.g., Equation 1 of \citealt{Levy2022}). This enables us to model the cluster surface brightness profile as a Gaussian and to determine the local background level (with other bright sources removed), which we subtract from the radial profile and subsequent fits. To obtain uncertainties on the fitted Gaussian, we perform a Monte Carlo simulation over the surface brightness uncertainties. Our uncertainties come from the standard deviation of 500 trials (shown as the blue shaded region in Figure \ref{fig:radial_profile}).

With this Gaussian model, we deconvolve the PSF in a simplified way. For this first analysis, we assume that the JWST PSF is Gaussian\footnote{We use the empirical PSF FWHM given in Table 1 of \url{https://jwst-docs.stsci.edu/jwst-near-infrared-camera/nircam-performance/nircam-point-spread-functions\#NIRCamPointSpreadFunctions-PSFFWHM}.}. Our estimate of the deconvolved (i.e., intrinsic) cluster radii is achieved by removing in quadrature the PSF half-width-half-maximum (HWHM) from the cluster Gaussian fitted HWHM. Through flux conservation, we calculate the deconvolved Gaussian profile and uncertainties, shown in red in Figure \ref{fig:radial_profile}.  We report these values in Table \ref{tab:catalog}.

Because of our approximate method to remove the PSF from our radius measurements, the deconvolved radii reported in Table \ref{tab:catalog} should be treated as a first estimate, not a robust measurement, of the intrinsic cluster radii. A more robust analysis is planned by our team (B. Cuevas-Otahola et al. in prep.) accounting for the actual PSF shape \citep[e.g., {\tt WebbPSF};][see also \citealt{Rigby2023}]{Perrin2014}.

\begin{deluxetable*}{ccccccccccc}
\tabletypesize{\scriptsize}
\tablecaption{NIRCam Catalog of Massive Star Cluster Candidates\label{tab:catalog}}
\tablehead{\colhead{ID} & \colhead{R.A.}    & \colhead{Decl.}   & \colhead{r$_\mathrm{F250M}$\,$^{a}$} & \colhead{F$_\mathrm{F250M}$\,$^{a}$} & \colhead{log(M$_\mathrm{*,0Myr}$)\,$^{b}$} & \colhead{log(M$_\mathrm{*,8Myr}$)\,$^{b}$} & \colhead{m$_\mathrm{F140M}$\,$^{c}$} & \colhead{m$_\mathrm{F250M}$\,$^{c}$} & \colhead{m$_\mathrm{F360M}$\,$^{c}$} & \colhead{Cross-Match\,$^{d}$}\\              & \colhead{(J2000)} & \colhead{(J2000)} & \colhead{(pc)}                              & \colhead{(mJy)}                        & \colhead{log(M$_\odot$)}                    & \colhead{log(M$_\odot$)}                    & \colhead{AB mag}                       & \colhead{AB mag}                       & \colhead{AB mag}                       &}
\startdata
0 & $9^{\mathrm{h}}55^{\mathrm{m}}46.651^{\mathrm{s}}$ & $+69^\circ40{}^\prime37.887{}^{\prime\prime}$ & 1.0$\pm$0.1 & 10.03$\pm$0.61 & 6.2$\pm$0.1 & 6.0$\pm$0.1 & 14.29 & 14.28 & 14.35 & 1, 2\\
1 & $9^{\mathrm{h}}55^{\mathrm{m}}50.437^{\mathrm{s}}$ & $+69^\circ40{}^\prime45.866{}^{\prime\prime}$ & 1.6$\pm$0.1 & 9.85$\pm$0.58 & 6.2$\pm$0.1 & 5.9$\pm$0.1 & 16.91 & 15.09 & 14.25 & 1\\
2 & $9^{\mathrm{h}}55^{\mathrm{m}}50.085^{\mathrm{s}}$ & $+69^\circ40{}^\prime45.938{}^{\prime\prime}$ & 1.4$\pm$0.1 & 5.50$\pm$0.34 & 6.0$\pm$0.1 & 5.7$\pm$0.1 & 16.25 & 15.43 & 14.82 & 1\\
3 & $9^{\mathrm{h}}55^{\mathrm{m}}50.848^{\mathrm{s}}$ & $+69^\circ40{}^\prime47.406{}^{\prime\prime}$ & 2.9$\pm$0.3 & 5.44$\pm$0.49 & 6.0$\pm$0.1 & 5.7$\pm$0.1 & 18.0 & 17.39 & 16.44 & 1\\
4 & $9^{\mathrm{h}}55^{\mathrm{m}}53.386^{\mathrm{s}}$ & $+69^\circ40{}^\prime50.480{}^{\prime\prime}$ & 2.1$\pm$0.1 & 5.28$\pm$0.28 & 6.0$\pm$0.1 & 5.7$\pm$0.1 & 16.67 & 16.23 & 15.75 & 1, 2\\
5 & $9^{\mathrm{h}}55^{\mathrm{m}}51.220^{\mathrm{s}}$ & $+69^\circ40{}^\prime47.666{}^{\prime\prime}$ & 1.4$\pm$0.1 & 5.22$\pm$0.46 & 6.0$\pm$0.1 & 5.7$\pm$0.1 & 16.1 & 15.64 & 15.01 & 1, 2\\
\enddata
\tablecomments{Table \ref{tab:catalog} is published in its entirety in the machine-readable format. A portion is shown here for guidance regarding its form and content.}\tablecomments{The NIRCam candidate star cluster catalog for M\,82. This table is ordered first by decreasing stellar mass then by decreasing flux. See Section \ref{sec:prop} (especially Section \ref{ssec:finalcatalog}) for details.}
\tablenotemark{\footnotesize a}{These properties are derived from the radial profile fitting described in Section \ref{ssec:radii}.}

\tablenotemark{\footnotesize b}{These stellar masses are derived as described in Section \ref{ssec:masses} for the ages given in the subscripts.}

\tablenotemark{\footnotesize c}{The magnitudes are calculated as described in Appendix \ref{app:apphot}.}

\tablenotemark{\footnotesize d}{Cluster candidates overlap with (1) \citet{McCrady2003} and/or (2) \citet{Mayya2008}.}

\end{deluxetable*}

\subsection{Stellar Mass Estimates}
\label{ssec:masses}

We measure the spectral flux density of each cluster from the area under the Gaussian fit to the F250M radial profile before deconvolution: $F = 2\pi I_{\rm peak} r_{\rm HWHM}^2$, where $F$ is the spectral flux density in mJy, $I_{\rm peak}$ is the peak intensity of the Gaussian profile, and $r_{\rm HWHM}$ is the cluster radius before PSF deconvolution\footnote{To recover $r_{\rm HWHM}$ before deconvolution, add the F250M PSF HWHM (0.75~pc) to the deconvolved values in Table~\ref{tab:catalog} in quadrature.}. We report these spetcral flux density values and propagated uncertainties in Table \ref{tab:catalog}. 

To convert the measured spectral flux densities into a stellar mass via a mass-to-light ratio (\ml), we must assume a cluster age, as the F250M data alone cannot constrain the cluster ages. 
For this first analysis, we take two age estimates as bounds to the true age of this cluster population. For a lower age limit, we assume a zero age main sequence (ZAMS) stellar population. This limit is applicable to the youngest, most embedded, cluster candidates in the burst. In the \yggdrasil\ SSP models, all of the stars start on the main-sequence at age~$=0$~Myr \footnote{The \yggdrasil\ models correct for pre-main sequence stars for $0.8\leq {\rm M_*/M_\odot} \leq 7$ \citep{Zackrisson2001}.}.

We assume 8~Myr as our upper age limit, following the analysis of \citet{Mayya2008} which built on primarily optical results from \citet{Satyapal1997}, \citet{ForsterSchreiber2003}, and \citet{Melo2005}. Because the NIR is less susceptible to dust extinction, we are able to probe down to more embedded, likely younger, clusters than optical studies. Older, more evolved clusters are visible in both optical and NIR observations. Therefore, the mean age of clusters probed with NIR measurements will be slightly younger than those detected in optical studies. Indeed, \citet{Satyapal1997} find an average age of 6~Myr for their NIR-identified star clusters in M\,82 compared to 8~Myr for the optically-identified clusters\footnote{\change{We note that while \citet{Satyapal1997} derived their ages using an instantaneous burst at solar metallicity, they assume a \citet{Salpeter1955} initial mass function and their stellar tracks do not include AGB stars. As AGB stars tend to have the most effect on the derived properties for ages $\gtrsim$10~Myr (Appendix~\ref{app:m2l_age}) such effects are not likely to change the ages estimated by \citet{Satyapal1997} drastically.}}. To be conservative, we take 8~Myr as a likely upper limit on the (representative) cluster age.

We determine the \ml\ using the \yggdrasil\ single stellar population (SSP) models \citep{Zackrisson2011}. This code (which has also been used to analyze NIRCam observations of star clusters in luminous infrared galaxies; \citealt{Linden2023}) computes the evolution of an instantaneous burst of star formation in a $10^6$~\msun\ cluster assuming a \citet{Kroupa2001} initial mass function and Padova-AGB stellar evolution tracks \citep[e.g.,][]{Bertelli2008,Bertelli2009}. We adopt models at solar metallicity, since the metallicity in the central 500~pc of M\,82 is roughly solar \citep{Lopez2020}. For simplicity, we assume maximal nebular emission ({\tt f$_{\rm cov}=1$}). We determine the luminosity of the \yggdrasil\ cluster in the F250M filter by convolving the spectral energy distribution (SED) from the \yggdrasil\ model with the F250M filter throughput (version 5.0, November 2022)\footnote{\url{https://jwst-docs.stsci.edu/jwst-near-infrared-camera/nircam-instrumentation/nircam-filters\#NIRCamFilters-filt_trans}}, yielding the luminosity of a $10^6$~\msun\ cluster in the F250M filter for each assumed age, which we then convert to the equivalent specrtral flux density. The resulting \ml\ in the F250M filter are 0.35~\msun/\lsun\ for 0~Myr and 0.18~\msun/\lsun\ for 8~Myr. At the distance of M\,82, these translate to $1.7\times10^5$~\msun~mJy$^{-1}$ and $8.8\times10^4$~\msun~mJy$^{-1}$ respectively. We multiply the measured spectral flux density (and uncertainties) of each cluster by \ml\ to obtain their stellar masses (\mstar) for each assumed age (Table~\ref{tab:catalog}). As we discuss in Appendix~\ref{app:m2l_age}, the ZAMS stellar masses correspond to the median for a $<10$~Myr star cluster population (where the older starburst in M\,82 occurred 10~Myr ago; \citealt{ForsterSchreiber2003}). The minimum \ml\ occurs at $\sim$8~Myr, and hence those stellar masses reflect the minimum cluster masses.


\subsection{The Final Candidate Massive Star Cluster Catalog}
\label{ssec:finalcatalog}

After the cluster identification, radius measurements, and stellar mass estimates, we implement the following cuts to build the final massive candidate star cluster catalog, presented in Table \ref{tab:catalog}. We note the number of candidates removed in parentheses at the end of each step below. 

\begin{enumerate}
    \itemsep0em
    \item Red supergiants (RSGs), asymptotic giant branch (AGB) stars, and other massive stars emit strongly at 2.5\micron\ and can be confused with massive star clusters based on their intensity alone \citep[e.g.,][]{Bolatto2007,Levesque2018,Boyer2024}. In order to remove these objects from our star cluster catalog, we measure the concentration of the star cluster from the radial profile to determine if the source is extended or consistent with a point source. We remove cluster candidates with $\geq$90\% of the flux concentrated in the central 2 pixels. (-99)
    \item Next, we remove candidates with bad radius fits, indicating a poor Gaussian fit which tends to produce solutions with unphysically large radii. \change{These correspond to weak clusters which are not well separated from the local background, resulting in a relatively flat profile and hence a poor Gaussian fit with a large radius. We remove clusters with deconvolved diameters {$>6$~pc}. When the images of candidates with diameters {$>6$~pc} are examined, these objects either appear as very small, weak sources in high background regions or highly extended and not cluster-like.} (-155)
    \item In order to present a catalog of massive star clusters, we implement a stellar mass cut of $10^4$~\msun\ assuming a ZAMS population. In addition, the distinction between star clusters and OB stars becomes ambiguous below $\sim10^3$~\msun\footnote{Based on the {\sc slug} models, which account for effects of stochastically sampling the IMF at low cluster masses, \citet{Krumholz2015} find that a $10^3$~\msun\ cluster has a bolometric luminosity of $5\pm4\times10^5$~\lsun. Individual O and (massive) B stars have bolometric luminosities which overlap with this range.}. Therefore, a lower mass limit of $10^4$~\msun\ places us solidly in the star cluster regime. (-154)
    \item Finally, we remove one candidate which overlaps with a known background galaxy (2MASS J09555095+6940302; \citealt{Skrutskie2006}) and one candidate which overlaps with the galactic center and appears highly elongated rather than cluster-like. (-2)
\end{enumerate}

\begin{figure}
    \centering
    \includegraphics[width=\columnwidth]{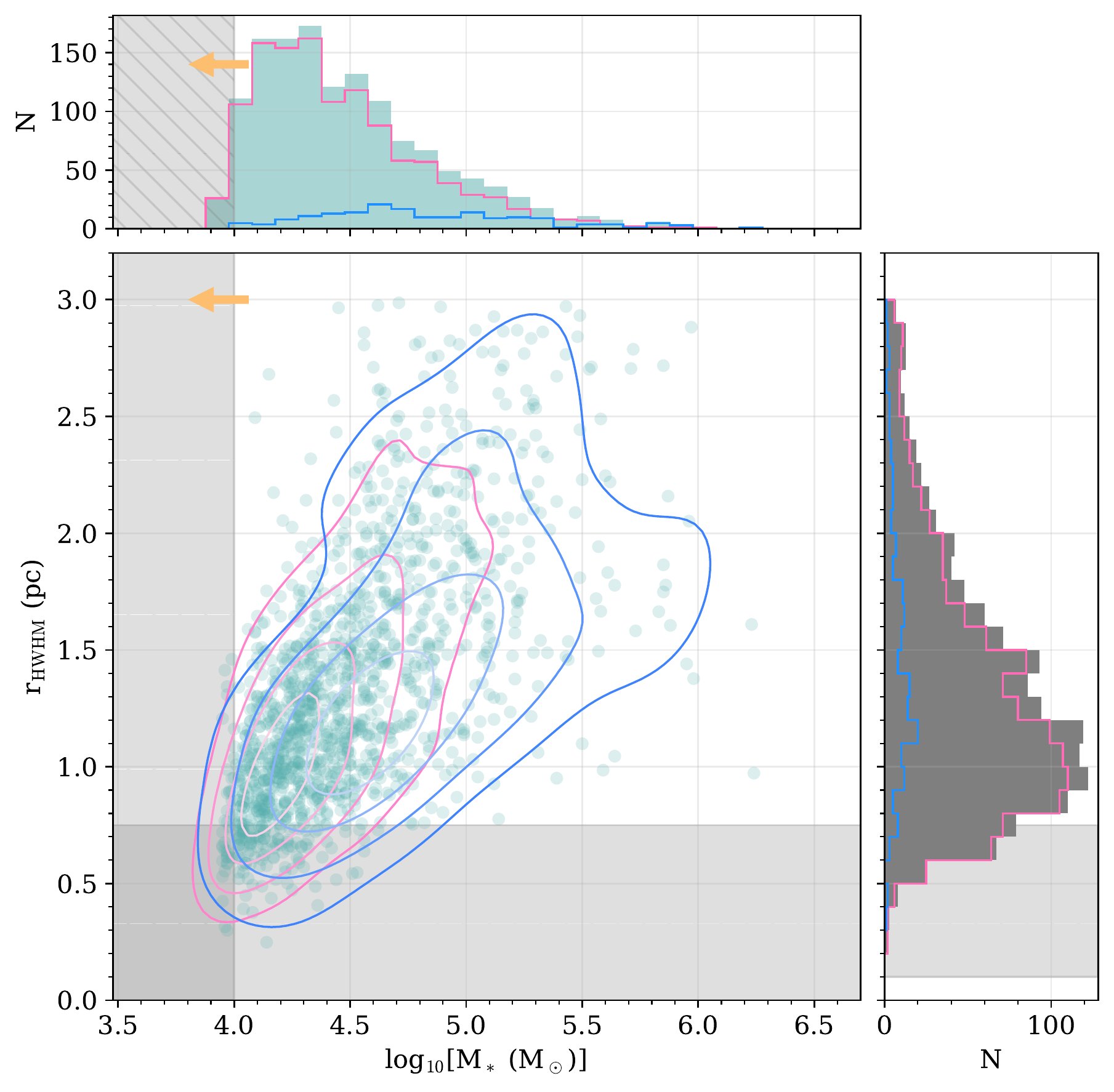}
    \caption{\change{(Bottom left) The mass-radius diagram of the final cluster catalog assuming age~$=0$~Myr. Typical uncertainties on both quantities are $\sim$10\%. Contours show the 10, 25, 50, and 75\% density levels for clusters which overlap with the \citet{Mayya2008} catalog and those only detected with NIRCam (pink). The horizontal gray shaded region shows r$_{\rm HWHM}$ less than the F250M HWHM PSF size. The vertical gray shaded region shows M$_*\leq10^4$~\msun. The orange arrow shows the shift assuming age~$=8$~Myr. (Right) A histogram of the deconvolved cluster radii given in Table~\ref{tab:catalog} (gray). The blue and pink histograms show the distribution for clusters which overlap with the \citet{Mayya2008} catalog and those which are only detected in NIRCam, respectively. The horizontal axis shows the number of clusters. The gray shaded region is the same as in the mass-radius diagram. (Top) A histogram of the stellar masses for age~$=0$~Myr given in Table~\ref{tab:catalog} (gray). The pink and blue histograms are as in the right panel. The orange arrow shows the shift assuming age~$=8$~Myr. The vertical axis shows the number of clusters. The gray shaded region is the same as in the mass-radius diagram.}}
    \label{fig:mass-radius}
\end{figure}

After these cuts, the catalog (Table \ref{tab:catalog}) consists of 1357 star cluster candidates. We show a mass-radius diagram, along with histograms of the masses and radii, of the final cluster catalog in Figure~\ref{fig:mass-radius}. Our catalog only captures massive (M$_*>10^4$~\msun) clusters which are bright and somewhat extended in the F250M band. Deeply embedded, very young, or very compact clusters may be missed by our selection criteria. Future observations and analysis from this JWST program will result in a more complete census of the embedded, young clusters.

\change{We investigate whether there are trends between the measured masses and radii and the deprojected galactocentric radius ($R_{\rm GC}$) of each star cluster candidate. To calculate $R_{\rm GC}$, we assume that the center of M\,82 is ${\rm 09^h55^m51.6^s\ +69^\circ40^\prime45.6^{\prime\prime}}$ (J2000) \citep{Bolatto2024} and that M\,82 has a major axis position angle of 67$^\circ$ \citep{Martini2018} and an inclination of 80$^\circ$ \citep{Lynds1963}. We find that there is no correlation between $R_{\rm GC}$ and cluster radius (Spearman rank correlation coefficient, $r_s$ = -0.05). There is, however, a weak negative correlation between $R_{\rm GC}$ and the stellar mass, with $r_s$ = -0.48. We note that $r_s$ = -0.41 if we assume M\,82 is perfectly edge-on. There is no significant difference in $r_s$ for star clusters which are detected with HST versus only detected with NIRCam.}

\citet{ForsterSchreiber2003} used NIR and MIR spectroscopy at 1.5\arcsec\ (26~pc) spatial resolution to build a model of the star formation history of M\,82. They find evidence for two bursts of star formation occurring 5~Myr and 10~Myr ago. From their models, the younger (older) burst produced $4.2\times10^7$~\msun\ ($1.6\times10^8$~\msun) of stars in the inner 500~pc, resulting in a total of $2\times10^8$~\msun\ of stars from both bursts (i.e., within the last 10~Myr). The total stellar mass of our star cluster candidates (assuming an age of 8~Myr to give a lower limit; see Figure~\ref{fig:mass-radius}) is $\approx4\times10^7$~\msun. This stellar mass is a factor of 4 lower than the stellar mass expected from the older burst model from \citet{ForsterSchreiber2003}. On the other hand, if we assume that all of our clusters have an age of 0~Myr, we find a total stellar mass of $\approx8\times10^7$~\msun. This stellar mass is a factor of 2 larger than the stellar mass expected from the younger burst model from \citet{ForsterSchreiber2003}. This points to the star clusters we identify in the center of M\,82 having a spread in age, though we cannot yet say whether they are consistent with finite bursts or a more continuous mode of star formation.

\section{Discussion}
\label{sec:disc}

\subsection{Comparison to other Star Cluster Catalogs}
\label{ssec:comp_other_cats}

A number of star cluster catalogs have been defined for M\,82. Here we focus on comparisons to two catalogs\footnote{We note that an SMA-based catalog of very young/embedded star clusters is forthcoming (\citepJD).}: 1) the HST- and Keck-based catalog of \citet{McCrady2003} and 2) the HST-based catalog of \citet{Mayya2008}. We refer the reader to those papers for details on observations and selection criteria. For both HST-based catalogs, we found it necessary to shift the positions to match the astrometry of the NIRCam data; these adjustments were done by-eye to match several of the brightest cluster candidates. We shifted the \citet{McCrady2003} positions by $(\alpha,\delta)=(-1.93^{\prime\prime},0.82^{\prime\prime})$ and the \cite{Mayya2008} positions by $(\alpha,\delta)=(-1.52^{\prime\prime},-0.67^{\prime\prime})$. Both of these catalogs cover regions that are larger than the field of view (FoV) of our NIRCam {\sc sub640} images. Within this FoV, the \citet{McCrady2003} catalog contains 19 star clusters and the \citet{Mayya2008} catalog contains 284 star clusters.

For a NIRCam star cluster candidate to be associted with a cluster in these previous catalogs, the coordinates of the cluster should be separated by less than the sum of the cluster radii in each catalog. For the NIRCam catalog, we use the radii listed in Table~\ref{tab:catalog}. For the \citet{McCrady2003} catalog, we use 0.6\arcsec\ (10.5~pc), which is the quoted precision of their astrometry. For the \citet{Mayya2008} catalog, we use a radius of 0.29\arcsec\ (5~pc), as shown in Figure~\ref{fig:multicolor}. Our cross-matching algorithm allows for multiple NIRCam sources to be matched to the same cluster in the other catalogs, and thus accounts for the source fragmentation due to our increased angular resolution.

Our cluster catalog recovers 100\% of of the stars clusters identified by \citet{McCrady2003} and 51\% of those identified by \citet{Mayya2008} (within the NIRCam {\sc sub640} FoV). Compared to the \citet{Mayya2008} HST catalog, we find 1183 previously unknown star cluster candidates. In other words, 87\% of the cluster candidates we identify with these NIRCam data are new. \change{As shown in Figure~\ref{fig:mass-radius}, the clusters that are newly detected with NIRCam tend to have smaller sizes and stellar masses than those that are also detected by \citet{Mayya2008} with HST.}

Differences between our NIRCam star cluster candidate catalog and that from HST \citep{Mayya2008} are due primarily to two effects. \change{First, that we only recover 51\% of the sources from the \citet{Mayya2008} catalog is primarily due to the increased angular resolution of JWST compared to HST. As is illustrated in the lower left panels of Figure~\ref{fig:multicolor}, there are many clusters identified with HST (light blue circles in the right panel) In this region of M\,82.} In the JWST image, however, many of the large, bright clusters seen with HST break apart into much smaller objects, which do not satisfy our star cluster selection criteria (Section~\ref{ssec:finalcatalog}). As a result, only 8 of the 26 clusters seen with HST in this region are classified as massive star cluster candidates here (blue circles). \change{Second, that we detect 1183 new star cluster candidates than previous HST catalogs is primarily due to dust extinction. As is illustrated in the lower middle and right panels of Figure~\ref{fig:multicolor}, we detect many star cluster candidates with NIRCam (pink circles) in these regions of M\,82 that are not present or classified as star clusters in the HST image and catalog.} We note that removing our stellar mass cut only increases the fraction of recovered HST clusters by 1\%.

\citet{McCrady2003} obtained Keck NIRSPEC spectroscopy on two massive star clusters in the center of M\,82, MGG-9 and MGG-11, allowing them to robustly measure their stellar masses and ages. Clusters MGG-9 and MGG-11 clusters correspond to our clusters 1 and 2, respectively, which are highlighted in Figure~\ref{fig:multicolor}. Using the HST NICMOS NIC2 F160W image, which has a pixel size of 1.3~pc (0.075\arcsec), \citet{McCrady2003} found half-light radii of 2.6~pc and 1.2~pc for MGG-9 and MGG-11, respectively. With our 1.8$\times$ smaller pixel scale for NIRCam data, we measure radii of 1.8~pc and 1.6~pc, respectively. By combining velocity dispersions measured from their high spectral resolution Keck NIRSPEC data with these radii, \citet{McCrady2003} report kinematic (virial) masses of $(1.5\pm0.3)\times10^6$~\msun\ and $(3.5\pm0.7)\times10^5$~\msun. They also estimate the ages of these clusters to be 10~Myr and 9~Myr respectively. Assuming these ages, we find stellar masses of $(0.9\pm0.8)\times10^6$~\msun\ and $(4.0\pm1.3)\times10^5$~\msun\ for clusters 1 and 2, respectively. We conclude that the masses we derive for these clusters agree with \citet{McCrady2003} within the uncertainties. 

\subsection{Color-Color Diagrams}
\label{ssec:colorcolor}

\begin{figure}
    \centering
    \includegraphics[width=\columnwidth]{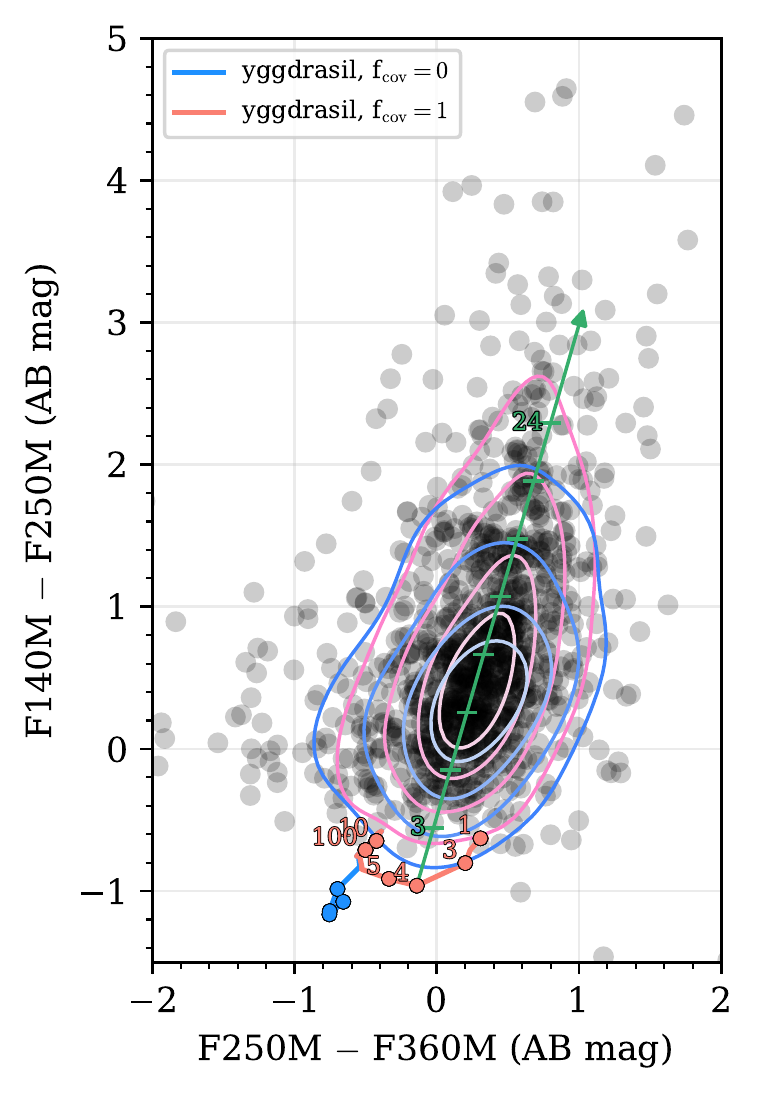}
    \caption{\change{A color-color plot of the clusters (black dots). The contours show the density of the points; contour levels show 10, 25, 50, and 75\% density levels for clusters which overlap with the \citet{Mayya2008} catalog (blue) and those only detected with NIRCam (pink). The colored tracks show \yggdrasil\ SSP models assuming no ($\rm{f_{cov}}=0$; blue) and maximal ($\rm{f_{cov}}=1$; red) contributions from nebular emission. The green arrow shows the slope of the reddening vector with A$_{\rm V}=3-24$~mag marked in increments of 3. The clusters in M\,82 detected with NIRCam show significant reddening with respect to the dust-free SSP tracks.}
    }
    \label{fig:color-color}
\end{figure}

To investigate the relative dust content in these candidate massive star clusters, we plot a color-color diagram using the F140M, F250M, and F360M filters. The F140M filter is most affected by dust extinction whereas we expect that dust continuum emission contributes to the F360M emission in addition to the stellar continuum\footnote{We note that the 3.4\micron\ aliphatic PAH emission feature and the 3.47\micron\ PAH plateau can also contribute to the F360M filter \citep{Sandstrom2023,Bolatto2024}. This will result in excess emission in the F360M filter that is not accounted for in the slope of the reddening vector, and hence our extinction values may be slightly underestimated. However, spectroscopic studies of star clusters in starburst galaxies find that these features are very weak when measured towards star clusters (S. Linden et al. in prep.).}. Thus, these colors can give us a handle on the relative dust content of these candidate star clusters, especially when compared to dust-free SSP models. 

We show the AB magnitudes derived from aperture photometry, described in Appendix~\ref{app:apphot}, in Figure~\ref{fig:color-color}. We also show tracks from the \yggdrasil\ \citep{Zackrisson2011} models for pure stellar emission ({\tt f$_{\rm cov}=0$}) and maximal nebular contribution ({\tt f$_{\rm cov}=1$}). These two tracks converge for ages $\gtrsim7$~Myr because, under the assumption of an instantaneous burst, most massive stars have died and are not replenished, so that the Lyman continuum flux is too low to sustain appreciable nebular emission. 

We plot the reddening vector in Figure~\ref{fig:color-color} following Equation~9 of \citet{Salim2020} for a range of A$_\mathrm{V}$. \change{The clusters which are only detected with NIRCam show somewhat larger extinctions compared to those also detected with HST by \citet{Mayya2008}, as expected (c.f. the pink and blue contours in Figure~\ref{fig:color-color}). In particular, the clusters which are detected with both HST and NIRCam have minimum extinctions $\approx$0~mag, whereas those that are only detected with NIRCam have minimum extinctions of $\approx$3~mag.} Overall, the cluster candidates show significant reddening, with A$_{\rm V}\sim3-24$~mag.  This range of extinctions agrees well with the extinctions measured in the NIR by \citet{McCrady2003}, who found A$_{\rm V}\approx11\ {\rm and}\ 7$ for MGG-9 and MGG-11, respectively (clusters 2 and 3 in our sample). For the optically-identified HST clusters, \citet{Mayya2008} report A$_{\rm V}\lesssim6$; we are able to probe more highly extincted clusters with these NIRCam data. Our range of A$_{\rm V}$ is also in excellent agreement with A$_{\rm V}=10\pm5$ derived by \citet{ForsterSchreiber2001} for a foreground screen model using multiple recombination lines in the integrated emission in the central region. In the NIRCam bands themselves, A$_{\rm V}\sim3-24$~mag corresponds to A$_{\rm 1.4\mu m}\sim0.7-5.4$~mag, A$_{\rm 2.5\mu m}\sim0.3-2.1$~mag, and A$_{\rm 3.6\mu m}\sim0.1-1.2$~mag.

The reddening vector in Figure~\ref{fig:color-color} is anchored to the {\tt f$_{\rm cov}=1$} SSP at an age of 4~Myr. For this choice, the reddening vector is approximately aligned to the major axis of the density contours. While there is a large spread in the potential cluster ages, this suggests that we are probing a relatively young, heavily reddened cluster population in the center of M\,82.

\subsection{Cluster Mass Function}
\label{ssec:cmf}

\begin{figure*}
    \centering
    \includegraphics[width=\textwidth]{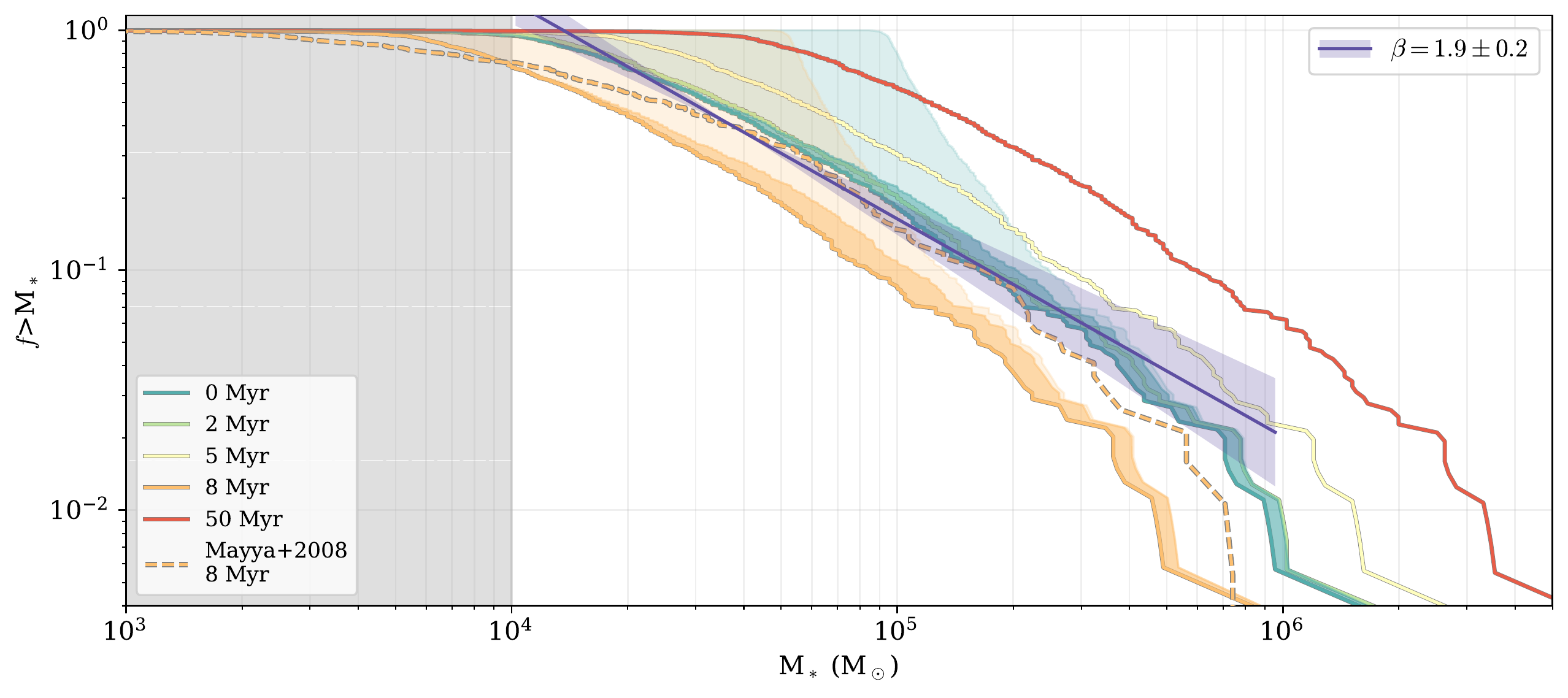}
    \caption{The cumulative CMFs shown as a fraction of the number of clusters for different assumed ages (colored solid curves). The blue and orange shaded regions on the 0~Myr and 8~Myr curves indicates the approximate range of mass increase due to extinction \change{which scales with M$_*$ (dark) or which is constant with M$_*$ (light); see Section~\ref{sssec:dust} for details.} The gray shaded region shows masses below $10^4$~\msun\ which are not included in the final catalog. The purple line and shaded region shows the power-law fit for the 0~Myr CMF for M$_*>10^4$~\msun. The dashed orange curve shows the cumulative CMF for the nuclear star clusters identified by \citet{Mayya2008} which have an assumed age of 8~Myr.}
    \label{fig:cmf}
\end{figure*}

The cumulative cluster mass function (CMF) measures the number of star clusters above a given stellar mass as a function of the stellar mass and may provide insights into the efficiency of star formation and links between the giant molecular cloud core mass function and the stellar initial mass function. CMFs are generally fit by a single power-law of the form $dN/dm\propto m^{-\beta}$, and observations generally measure an index $\beta\approx2.0\pm0.2$ \citep[e.g.,][]{Zhang1999,Mayya2008,Krumholz2019,Emig2020,Mok2020,Whitmore2021}. However, some studies suggest that there may be a cutoff of the power-law at the high-mass end \citep[e.g.,][]{Larsen2009,Konstantopoulos2013,Adamo2015,Messa2018}, potentially indicating that there is some maximum stable cluster mass above which proto-clusters fragment, though other studies find that such cutoffs are statistical rather than physical \citep[e.g.,][]{Cook2019,Mok2019,Mok2020,Whitmore2020}. \change{While most studies of star clusters in other galaxies are cataloged using optical data \citep[e.g., broadband HST photometry;][]{Zhang1999,Mayya2008,Mok2020,Whitmore2020}, the CMF slopes are consistent even when high-resolution submillimeter data are used, probing the most embedded, youngest star cluster populations \citep[e.g.,][]{Emig2020}.} \change{Given the resolution, sensitivity, and range of extinctions probed by these JWST NIRCam data}, we are in a prime position to investigate the shape of the CMF in the extreme starburst of M\,82.

We construct the cumulative CMF for various assumed ages for the NIRCam-detected clusters. We follow the methodology described in Section \ref{ssec:masses} to calculate the stellar masses for 0, 2, 5, 8, and 50~Myr. We plot these cumulative CMFs (in terms of the fraction of clusters, rather than the absolute number) in Figure \ref{fig:cmf}. We discuss the progression of the CMFs (i.e., \ml) with age in Appendix~\ref{app:m2l_age}.

The cumulative CMFs exhibit significant flattening at the low-mass end, which results from our choice to remove clusters with a ZAMS M$_*<10^4$~\msun. All of the cumulative CMFs show a steep turnover at the high-mass end, which is commonly seen in other galaxies\change{, using both optical \citep[e.g.,][]{Mok2020} and submillimeter \citep[e.g.,][]{Emig2020} observations proving a range of cluster ages and embeddedness.}

We compare our 8~Myr CMF to the CMF derived for the ``nuclear" (i.e., central, non-disk) clusters identified with HST by \citet{Mayya2008}; their cumulative CMF is shown in the orange dashed curve in Figure~\ref{fig:cmf}. Overall the cumulative CMFs at 8~Myr show some differences. At the low-mass end, we recover slightly more clusters compared to \citet{Mayya2008}, likely due to the increased resolution and sensitivity of JWST. At the high-mass end, our cumulative CMF drops off faster than \citet{Mayya2008}, likely because large clusters break apart at this increased spatial resolution compared to HST.

We fit our CMFs where M$_*>10^4$~\msun\ with a power-law of the form $\log y\propto -\alpha\log x$, where $\beta = 1+\alpha$. For a range of ages, we find $\beta=1.9\pm0.2$ (where the value is the slope of the ZAMS CMF and the uncertainty reflects the range of slopes measured for different ages), as shown in Figure \ref{fig:cmf}. This is in excellent agreement with CMF slope measurements in other starburst galaxies, which tend to have $\beta\approx2$ \citep[e.g.,][]{Emig2020,Mok2020,Whitmore2021}, and with the HST-based star cluster catalog of \citet{Mayya2008} who found $\beta=1.8\pm0.1$. However, from Figure~\ref{fig:cmf} it is clear that a simple power-law is not a particularly accurate description of the CMF shape. 



\subsubsection{Effects of Dust Extinction}
\label{sssec:dust}

Because of dust, the stellar mass estimates calculated in Section~\ref{ssec:masses} are likely underestimated since the SSP models used to derive the mass-to-light ratio do not account for the effects of dust extinction or emission. For the mean F250M AB magnitude ($20\pm1$~mag), A$_{\rm 2.5\mu m}=0.3-2.1$~mag (Section~\ref{ssec:colorcolor}) corresponds to a spectral flux density increase of $<0.5$~mJy or a stellar mass increase of $<8\times10^4$~\msun\ ($<4.7\times10^4$~\msun) for a ZAMS (8~Myr) population. We note, however, that this change will not result in a constant shift for all masses and, therefore, could impact the measured CMF slope. The precise change in mass depends on the extinction of each individual cluster which itself likely depends on the cluster mass \citep[e.g.,][]{deMeulenaer2013} \change{as well as the cluster's exact age}. \change{With future spectroscopy, the age and reddening can be constrained individually, leading to much more robust mass measurements.}

\change{As the most pessimistic case, we also test the change in the CMF slope assuming a constant A$_{\rm 2.5\mu m}=2.1$~mag across all masses. We show the potential shift in the 0~Myr and 8~Myr cumulative CMFs as the blue and orange light shaded regions in Figure~\ref{fig:cmf}. This pessimistic case will result in a steepening of the CMF slope up to $\beta=3.2$. Such a steep CMF slope is inconsistent with literature-reported values. }

\change{As a more realistic way to quantify the effect of extinction on the shape of the CMF, we assume that the most massive clusters are the most extincted \citep[e.g.,][]{deMeulenaer2013}. We implement this a simple linear scaling of A$_{\rm 2.5\mu m}$, ranging from 0.3~mag at the lowest mass to 2.1~mag at the highest mass. We show the potential shift in the 0~Myr and 8~Myr cumulative CMFs as the blue and orange dark shaded regions in Figure~\ref{fig:cmf}. For such a change to the CMF, $\beta=1.8$, within the quoted error.}

\section{Summary}
\label{sec:summary}
JWST has opened a new window into our understanding of star cluster formation and evolution. In this letter, we use the first NIRCam images of the prototypical starburst galaxy M\,82 \citep{Bolatto2024} to catalog and analyze the NIR-emitting star clusters in its center.

\begin{enumerate}
    \itemsep0em
    \item We identify 1357 star cluster candidates with M$_*>10^4$~\msun\ in the nuclear starburst of M\,82 based on the NIRCam F250M images (Table~\ref{tab:catalog}; Figure~\ref{fig:multicolor}). Compared to the HST star cluster catalog of \citet{Mayya2008}, we find overlap with $\approx$50\% of their sample and identify 1183 previously undetected star cluster candidates.
    \item Based on the color-color diagrams and comparing to dust-free \yggdrasil\ SSP models, we find that the star cluster candidates we detected with NIRCam still exhibit heavy dust extinction, with A$_{\rm V}\approx3-24$~mag, corresponding to A$_{\rm 2.5\mu m}\sim0.3-2.1$~mag (Figure~\ref{fig:color-color}).
    \item We estimate stellar masses for the star cluster candidates based on the \yggdrasil\ stellar population synthesis models and construct the CMF (Figure~\ref{fig:cmf}). We find a power-law CMF slope of $\beta=1.9\pm0.2$, in excellent agreement with studies of star clusters in other starburst galaxies \citep[e.g.,][]{Emig2020,Mok2020}.

\end{enumerate}

Spectroscopy of these clusters in the NIR and MIR (e.g., with NIRSpec and MIRI MRS onboard JWST) will enable us to measure robust ages \change{and extinctions} for the star clusters, which is currently a key source of uncertainty in our mass measurement. In the future, MIRI imaging and spectroscopy data from this JWST GO program \#1701 will enable us to find younger and more embedded star clusters than is possible with the F250M NIRCam filter alone. By cataloging the clusters identified with HST \citep[e.g.,][]{McCrady2003,Mayya2008}, NIRCam, MIRI, and the SMA (e.g., \citepJD), we will have a complete census of the star cluster population across evolutionary stages in the prototypical starburst in M\,82. 

\begin{acknowledgments}
R.C.L. thanks Erik Zackrisson for producing the \yggdrasil\ model magnitudes for the NIRCam filters used here. R.C.L. also thanks Sean Linden and the PHANGS-JWST/HST Stellar Populations Working Group for helpful discussions and advice.
R.C.L. acknowledges support for this work provided by a National Science Foundation (NSF) Astronomy and Astrophysics Postdoctoral Fellowship under award AST-2102625. A.D.B. and S.A.C. acknowledge support from the NSF under award AST-2108140. L.L. acknowledges that a portion of their research was carried out at the Jet Propulsion Laboratory, California Institute of Technology, under a contract with the National Aeronautics and Space Administration (80NM0018D0004). R.S.K. acknowledges financial support from the European Research Council via the ERC Synergy Grant ``ECOGAL'' (project ID 855130),  from the German Excellence Strategy via the Heidelberg Cluster of Excellence (EXC 2181 - 390900948) ``STRUCTURES'', and from the German Ministry for Economic Affairs and Climate Action in project ``MAINN'' (funding ID 50OO2206). V.V. acknowledges support from the ALMA-ANID Postdoctoral Fellowship under the award ASTRO21-0062. M.R. acknowledges support from project PID2020-114414GB-100, financed by MCIN/AEI/10.13039/501100011033. M.J.J.D. acknowledges support from the Spanish grant PID2022-138560NB-I00, funded by MCIN/AEI/10.13039/501100011033/FEDER, EU. This work is based on observations made with the NASA/ESA/CSA JWST. The data were obtained from the Mikulski Archive for Space Telescopes at the Space Telescope Science Institute, which is operated by the Association of Universities for Research in Astronomy, Inc., under NASA contract NAS 5-03127 for JWST. These observations are associated with program JWST-GO-01701. Support for program JWST-GO-01701 is provided by NASA through a grant from the Space Telescope Science Institute, which is operated by the Association of Universities for Research in Astronomy, Inc., under NASA contract NAS 5-03127. This research made use of \sourcextractor\footnote{https://github.com/astrorama/SourceXtractorPlusPlus}, an open source software package developed for the Euclid satellite project. This research has made use of NASA's Astrophysics Data System Bibliographic Services. 

\end{acknowledgments}

\vspace{5mm}
\facilities{JWST(NIRCam)}
\software{Astropy \citep{Astropy2022}, MatPlotLib \citep{matplotlib}, \texttt{multicolorfits} \citep{multicolorfits}, NumPy \citep{numpy}, pandas \citep{pandas}, photutils \citep{photutils}, SciPy \citep{scipy}, seaborn \citep{seaborn}, \sourcextractor\ \citep{sourcextractor++1,sourcextractor++2}, \yggdrasil\ \citep{Zackrisson2011}}

\bibliographystyle{aasjournal}

\appendix

\section{$\Upsilon_*$ as a function of Age}
\label{app:m2l_age}

\begin{figure*}
    \centering
    \includegraphics[width=\textwidth]{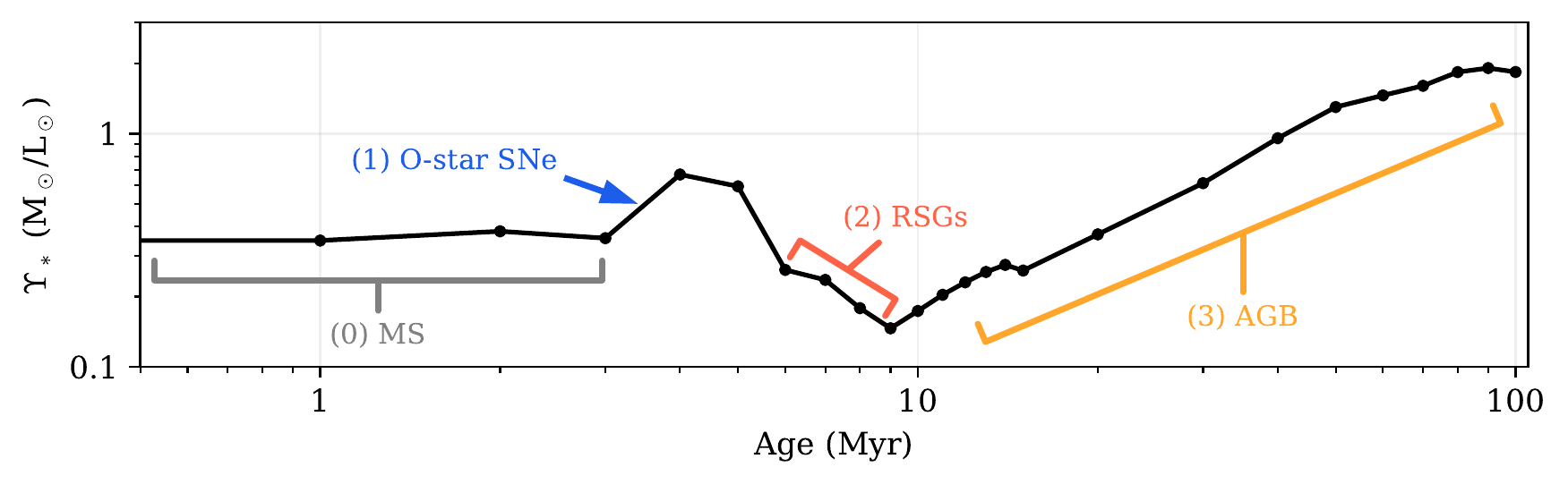}
    \caption{The progression of \ml\ in the F250M filter with age from the \yggdrasil\ models at solar metallicity and maximal nebular emission (see Section~\ref{ssec:masses}). Key phases of stellar evolution, which govern changes in \ml, are marked. See the text in Appendix~\ref{app:m2l_age} for details.}
    \label{fig:m2l_age}
\end{figure*}

As shown in Figure~\ref{fig:cmf}, the cumulative CMFs do not shift monotonically with age. This is due to how \ml\ varies with age in the \yggdrasil\ models, shown in Figure~\ref{fig:m2l_age} for the first 100~Myr, which is governed by stellar evolution. Here we present a broad-brush and highly simplified interpretation of the behavior of \ml\ with age in the F250M filter, where the numbers corresponds to those in Figure~\ref{fig:m2l_age}.
\renewcommand\labelenumi{(\theenumi)}
\begin{enumerate}
    \setcounter{enumi}{-1}
    \itemsep0em
    \item For the first $\sim$3~Myr of the cluster's evolution, \ml\ is roughly constant as stars age on the main-sequence (MS).
    \item After $\sim$3~Myr, the most-massive and most-luminous O-stars will rapidly evolve off the MS and explode as supernovae (SN). As these stars die, the luminosity of the cluster decreases resulting in an increase in \ml. Although the stellar mass of the cluster also decreases, the OB-stars dominate the cluster luminosity, so the decrease in luminosity is the dominant effect.
    \item From $\sim5-9$~Myr, \ml\ decreases. During this time, less massive O and B stars are evolving off the MS. While the supergiant phases have roughly constant bolometric luminosity, the stars become redder in color as they cool and expand. This increases the luminosity in the F250M band. The red supergiant (RSG) phase plays a key role here, as the bolometric luminosity, and hence F250M, increases \citep[see also section 4a of][]{Charlot1991}. Together, these effects significantly decrease \ml.
    \item Between $\sim9-100$~Myr, the luminosity in the F250M filter increases primarily due to the presence of increasing number of asymptotic giant branch (AGB) stars \citep[see also section 4a of][]{Charlot1991}. This increased luminosity decreases \ml.
\end{enumerate}
In the first 10~Myr (corresponding to the age of the older starburst in M\,82; \citealt{ForsterSchreiber2003}), the median \ml~$=0.35$~\msun/\lsun, equivalent to \ml\ for a ZAMS. Therefore, the stellar masses we compute for a ZAMS are likely the best reflection of the stellar mass distribution of the star clusters in the center of M\,82 when assuming a single \ml. The stellar masses derived for an 8~Myr population, on the other hand, are near the minimum possible masses, as they correspond to the minimum \ml. Therefore, the stellar masses reported in Table~\ref{tab:catalog} reflect the median (0~Myr) and minimum (8~Myr) stellar masses for the star clusters in the center of M\,82.

\section{Aperture Photometry}
\label{app:apphot}

To construct the color-color diagram presented in Section~\ref{ssec:colorcolor} and Figure~\ref{fig:color-color}, we measure the cluster brightness in the F140M, F250M, and F360M images in apertures (as opposed to the flux based on the radial profile fitting described in Section~\ref{ssec:radii} and reported in Table~\ref{tab:catalog}). We place circular apertures at the cluster positions in Table~\ref{tab:catalog}. Our aperture radii correspond to the 70\% encircled energy of the PSF ({\tt r$_{\rm 70}$}) given in the NIRCam APCORR Reference File\footnote{This file was accessed at \url{https://data.science.stsci.edu/redirect/JWST/jwst-data_analysis_tools/stellar_photometry/aperture_correction_table.txt} on 2024-03-22.}. For the F140M, F250M, and F360M filters {\tt r$_{\rm 70}$} corresponds to 0.065\arcsec, 0.092\arcsec, and 0.123\arcsec\ respectively. After extracting the flux in each aperture, we apply the corresponding aperture correction in the NIRCam APCORR Reference File ({\tt apcorr70}); for all three filters, {\tt apcorr70}~$\approx1.46$. We note that these aperture corrections are measured for point sources. For our clusters, which are slightly extended compared to the PSF, these point-source apertures corrections are not strictly correct. However, for the purposes of this preliminary investigation, we will proceed with these corrections.

We extract the background in an annulus around and with the same area as the circular aperture. We measure the median (50\%), 16\%, and 84\% background levels in the annulus. We use the median for the background subtraction and propagate the 16\% and 84\% values into the uncertainties.

Finally, we convert the flux extracted in each aperture and in each filter to AB magnitudes\footnote{The AB magnitudes are calculated following \url{https://jwst-docs.stsci.edu/jwst-near-infrared-camera/nircam-performance/nircam-absolute-flux-calibration-and-zeropoints\#NIRCamAbsoluteFluxCalibrationandZeropoints-ABmagnitudes}.} which we list in Table~\ref{tab:catalog}. The color-color diagram for the clusters following this procedure is shown in Figure~\ref{fig:color-color}.

\end{document}